# Arctic Connectivity: A frugal approach to infrastructural development

Abildgaard, Mette Simonsen
Aalborg University
Department of Culture and Learning
AC Meyers Vænge 15, 2450 Copenhagen SV, Denmark

Ren, Carina
Aalborg University
Department of Culture and Learning
AC Meyers Vænge 15, 2450 Copenhagen SV, Denmark

Leyva-Mayorga, Israel
Aalborg University
Department of Electronic Systems
Fredrik Bajers Vej 7, 9220 Aalborg OE, Denmark

Stefanovic, Cedomir
Aalborg University
Department of Electronic Systems
AC Meyers Vænge 15, 2450 Copenhagen SV, Denmark

Soret, Beatriz
Aalborg University
Department of Electronic Systems
Fredrik Bajers Vej 7, 9220 Aalborg OE, Denmark

&

Popovski, Petar
Aalborg University
Department of Electronic Systems
Fredrik Bajers Vej 7, 9220 Aalborg OE, Denmark

**Abstract**

As the Arctic is heating up, so are efforts to strengthen connectivity within the region, but also to enhance the connections from remote settlements to the global networks of trade as well as sociality. With global interest in the Arctic on the rise, it becomes increasingly relevant to ensure that investments in arctic infrastructure actually serve the people of the Arctic, while promoting industrial and commercial innovation in the region through widespread access to broadband and Internet of Things (IoT) services. This calls for interdisciplinary research strategies that are able to connect and integrate technological and societal approaches, which are commonly applied separately and in isolation from one another. In this article, we propose an interdisciplinary collaborative research agenda for Arctic connectivity. Drawing on examples from Greenland, we stress the need for localized knowledge to design valuable and cost-effective connectivity solutions that cover the needs for everyday life and may also provide a new set of collaborative connectivity tools for innovation at an international level. Such solutions, termed 'frugal connectivity', are vital for the development of connected Arctic communities.

**Introduction**

As the melting of the ice makes way for new sailing routes and eases access to rare minerals and oil, the Arctic has gradually become the center of strategic geo-political attention. This has sparked an interest to link this remote region to a global network of trade and sociality and spurred political initiatives to strengthen "Arctic connectivity" through massive investments in infrastructure (Taksøe-Jensen, 2016; Arctic Council, 2017).

In academia, several fields have addressed issues relating to Arctic connectivity. In media and communication studies, notions such as 'center-periphery problems', the 'digital divide', and 'information poverty' (Rygaard, 2017; Drori, 2010; Subramony, 2007; Norris, 2001) emphasize the negative consequences of global and regional uneven use of ICT in developing and remote Arctic areas and point to a divide between the 'haves' and the 'have-nots' reinforced by infrastructure planning (Warschauer, 2002). Similarly, scholars examining the transformations in Arctic infrastructures tend to frame "the entire North, as a global periphery, [...] in relation to, and as a poorer version of, the South: terrible internet connections, bad roads, no services (Exner-Pirot, 2017:2)".

While recent years have shown a growing interest in the Arctic, moving it gradually towards a perceived geopolitical center stage (Bjørst & Ren, 2015), prevailing academic literature on communication and infrastructures, along with policy reports and planning efforts, still overwhelmingly approach the Arctic as remote and marginal, as disconnected. This is problematic not only because Arctic people have always been connected - travelling and exchanging information by umiaq (large boat), kayak, dog sledge, and other means - and for long, they did not think of themselves in any other way. But it is also problematic because it is based on a notion of connectivity which is underpinned by a universal logic of 'the more the better'. Knowledge is thus acutely lacking on how to grasp and develop Arctic connectivity through a more situated approach. To bridge this gap, we propose an interdisciplinary and collaborative research agenda drawing on Science and Technology Studies (STS) to explore *how connectivity is imagined and constructed as an everyday assemblage of humans, technologies, discourses and policies* in the context of physical and digital infrastructures.

To explicate this research agenda, we draw examples from Greenland, a territorially large, but population-wise small Arctic nation. The Greenlandic case is intended as a starting point. We are here not able to encompass the potential for frugal connectivity development as it relates to the entire Arctic, but will consider some contrasts and similarities throughout the article.

In Greenland, the establishment of seamless communication infrastructure has been a consistent endeavor since the early 20th century, where Greenland's first wireless radio telegraph station, the short-lived "Myggbukta", was established in 1922 by the Norwegian Meteorological Institute. At current, Greenlandic discourses on and efforts toward strengthening connectivity center around increasing independence, where strengthened social cohesion, business development, improved educational and medical services are mentioned as possible outcomes of facilitating more connectivity (see the national digital strategy for Greenland, Naalakkersuisut, 2018). This underlines how technological innovations and societal imaginaries are closely aligned (Jasanoff, 2015) and, as we propose here, should be studied accordingly.

Recent innovations in ICT (information and communication technology) offer promising avenues for Greenland's plans to achieve increased independence, to increase its participation in global value chains, and to offset geographical "disadvantages" and compensate for geographical separation through connectivity (WBG, 2019). New connectivity types, services, and applications are emerging all around the globe with the roll-out of the fifth generation of mobile networks (5G). The latter has, for some years now, been considered a disruptive digital technology, capable of promoting inclusion, efficiency and innovation (WBG, 2016). Differently from previous generations, 4G and 3G, 5G is designed to provide exceptionally high guarantees for the rapid delivery of data between not just humans, but also 'things', such as machines, sensors, robots, cars, or similar. In particular, 5G provides a platform for industrial innovation that targets the niche or specialized 'vertical industries', such as health, automotive, energy, and entertainment (5GPPP, 2015). Here the term 'vertical' stands as a metaphor that observes 5G as a connectivity platform, on top of which one vertically builds digital solutions for diverse industries.

While numerous communities across the globe are racing to deploy their 5G infrastructure, the roadmap for 5G in Greenland is quite different. *First*, the currently deployed infrastructure is able to provide private internet access and 4G coverage to more than 90% of the population. However, there are radical differences between densely and sparsely populated areas. The connectivity services exhibit exceedingly high costs outside urban areas, and they experience different performance in terms of data speed, latency (e.g., the time, usually expressed in milliseconds, it takes for the network to respond to a request made in an online game) and the other performance indicators of the connectivity services, collectively denoted as Quality of Service (QoS). Therefore, rushing to deploy 5G infrastructure in densely populated areas such as in the capital, Nuuk, may exacerbate the disparity between communities. *Second*, the long distance between settlements and the specific geographical conditions of the region makes the deployment of new infrastructure a costly and complicated task. In its current phase of standardization, 5G lacks features to connect remote communities more effectively than 4G. For this reason, other technologies, which are cost-effective, must be integrated. *Third,* as confirmed with ICT industry stakeholders in Greenland, the futuristic applications and performance-oriented goals of 5G have created a generalized lack of enthusiasm in Greenland towards it (Tele Greenland, pers. comm. 2020).

Reigniting the interest in novel technologies and securing their development in line with the needs of local communities and businesses calls for interdisciplinary research strategies that are able to connect and integrate societal and technological approaches that are usually studied independently. In particular, we consider that deploying infrastructure to guarantee the availability of broadband and IoT services in the Greenlandic society and industry, even as a slightly downscaled version, is essential to maintain international competitiveness, but also and in connection to this, to maintain and strengthen the creation of jobs and services outside of the main towns.

For this purpose, we introduce the principle of 'frugality': A careful balance of local needs and technological possibilities to make the most of existing resources while providing valuable and flexible solutions. In introducing frugal connectivity development, we aim to emphasize infrastructural development that follows everyday needs and fosters sustainable industrial and societal innovation in line with Sustainable Development Goals (SDGs) (UN, 2015). This requires deep knowledge of the local everyday and industrial practices and needs to provide valuable and flexible solutions which, in turn, nurture future infrastructure research and development. The principle of frugality introduced here draws on on-going research on frugal and rural connectivity. However, while studies of frugal and rural connectivity often calls for bottom-up approaches that are tailored to meet local needs, the existing literature is primarily technical in scope, lacking considerations of how such a bottom-up development approach *can* take local needs into account (Khaturia, Chaporkar and Karandikar, 2017; Quadri et al, 2011; Simba et al, 2011; Dhanajay et al, 2011; Yacooub and Alouini, 2020). A small number of anthropological studies of telecommunication policy and rural areas do exist (Gregg, 2010; Gregg and Bell, 2008), but they are isolated from the technical literature mentioned above. As a example, the predominant use of 'frugal' in relation to ICT is in reference to technical macro level planning, such as the 'frugal 5G initiative' (Khaturia, Jha and Karandikar, 2020), although rare examples also consider frugality as a local micro practice of 'making do' with available means (Lanerolle, 2018). We emphasise the need to encompass both levels of frugality to plan on the basis of local practices, and the need - explicated by the current gap in the literature - to study connectivity in an interdisciplinary approach integrating collaboration with local stakeholders, deep ethnographic studies of local connectivity practices, and technical expertise.

To approach and contextualize the concept of frugality within Arctic connectivity, we first present and discuss traditional approaches to connecting the Arctic and outline the essential aspects that must be covered by frugal approaches. Next, we analyze and discuss a case of Arctic digital connectivity, drawing on material from online sociability in Greenland. We then describe some distinctive characteristics and potential applications of IoT connectivity in Greenlandic industries. Based on this situated account of Arctic everyday connectivity and on the industrial landscape, we argue for ways of developing Arctic digital connectivity through a frugal approach, which supports such digital practices. Finally, we propose a future collaborative research agenda for a frugal approach to arctic connectivity that seeks to pragmatically tinker with, rather than fix or turn over the digital infrastructures, that currently shape and interfere with online Arctic sociality.

**Approaches towards frugal Arctic connectivity**

As Schweitzer and Povoroznyuk describe, a dominant infrastructural logic of the built environment in the Arctic is "mastering the north", a conceptualization of infrastructure as a means to modernize and "overcome remoteness" (2019, p 236) as well as gaining access to the resources that are sought after on a national or global, rather than a local, level. Schweitzer et al. primarily draw examples from transport infrastructures in the Russian Arctic (Schweitzer and Povoroznyuk, 2019; Schweitzer, Povoroznyuk and Schiesser, 2017), but in a Greenlandic context we can find similar examples of this infrastructural paradigm in the Danish administration's modernization plans for Greenland made in the 1950's and 60's (Grønlandskommissionen, 1950; Grønlandsudvalget af 1960, 1964). Here, the Greenlandic population was encouraged to move (often with few alternatives) from rural settlements to towns, where infrastructural development was prioritized.

This development exemplifies a top-down approach, where logics from elsewhere on how habitation in a modern nation should be designed were imported to project the future Greenlandic society. In spite of massive critique of these modernist development frameworks (Hvidtfelt Nielsen & Kjærgaard, 2016), similar development models thrive in contemporary discussions around Greenland's digital infrastructure. In such discussions, communication connections are evaluated using generic scales, leaving little room for tailoring or adaptation to the vast distances, scattered populations and severe weather that characterizes Greenland. An additional complicating factor is that the foundational tele-infrastructures were developed by the Danish administration through organs such as GTO (Grønlands Tekniske Organisation/Greenland's Technical Organization), establishing rationale and priorities, such as concentrating infrastructure and population in larger cities, whose long-lasting impacts on infrastructural development in Greenland have not yet been studied.

Such policies have reached a tipping point with respect to the rollout of the 5G system. On the one hand, the 'futuristic' use cases and greatly generic (promised) performance goals of 5G: hundreds of Mbps, millisecond latencies, 99.999% of reliability, and up to tens of thousands of user requests served per second create a deep sense of disconnection with the actual and potential needs of the population and industry sectors. While this sense of disconnection is slightly diminished by the focus of 5G in vertical industries, the real-life applications and benefits of 5G, even in self-proclaimed 'smart-cities,' are not entirely clear. It is important to note that 5G has an extremely high deployment cost that, in combination with the emphasis on network densification (that is, installation of more base stations per unit area), is currently prohibitively high for Greenland, even for the biggest operator Tele Greenland. Besides, the current infrastructure in Greenland provides widespread 4G coverage, serving over 90% of the population (WBG, 2019).

Regardless of the excess or lack of enthusiasm towards 5G, there is no denying that it has the potential to become a disruptive technology that will open the door for innovation in different sectors of societies around the globe. This is because 5G supports IoT connectivity as a 'native' feature, in the sense that IoT connectivity is not an add-on to a broadband service in 5G, but something that is part of the system from its initial conceptualization. Therefore, frugal approaches could provide the Arctic communities with 1) adequate broadband services across all communities, which are competitive in price and QoS to international standards; 2) a subset of essential IoT connectivity features with the same QoS guarantees as in other communities; 3) a downscaled, custom-tailored version of a subset of IoT connectivity features relevant for Arctic; and 4) new applications of IoT connectivity tailored for the Greenlandic communities.

In order to offer the reader a view into the current everyday as well as possible futures of Arctic connectivity, we will now discuss some examples from the small-scale and mundane to the more encompassing industrial uses and applications of the current communication infrastructures in contemporary Greenland. Our aim is to illustrate how future frugal approaches can work from a situated perspective and learn from the particularities of local connectivity enablers and disablers to tinker with, rather than revolutionize, Arctic infrastructure.

**Connectivity in the Arctic - infrastructures and their everyday use**

So far, research on social media in the Arctic and Greenland in particular has concentrated on SoMe as a platform for political mobilization. A study by Jørgensen on a protest movement against the new Greenlandic Parliament building that was mobilized on Facebook is a case that shows the *internal* political significance of social media in Greenland (2017). Another example is Yunes, who analyzed the #sealfie movement, which protests the negative portrayal of seal hunting by environmental celebrities such as Ellen DeGeneres, to show how Arctic online mobilization also has political impact *outside* the region's limits (2016). However, in their exploration of how current representations of the Arctic are (re)produced or challenged online, Ren & Munk (2019) argue that studies on the ground are still lacking, for instance on how the Arctic is represented on social media platforms. This points to a gap in understanding the relationships and impacts of connectivity to online sociality and everyday connectivity in an Arctic context.

So how can we think of other, more situated ways in which to understand and develop Arctic connectivity, in particular when considering communication infrastructures, which are inherently heavy and cost-consuming? An alternate logic can be developed by taking inspiration from already existing mundane communication practices: By looking at and learning from the everyday use practices of existing technologies in a locally adapted and mundane way, it is, as we shall later see, one which draws on frugality and making-do.

Greenland has an estimated number of 40 000 users on Facebook. In comparison, 14 000 are on Instagram, and 3 600 on Twitter. This means that a striking 86% percent of the population over 13 years of age has a profile on Facebook (Hootsuite/We are social, 2019). Another particularity about the use of Facebook in Greenland is that people are more active than in most other nations in the world, especially in relation to sharing, commenting and liking posts (http://greenlandtoday.com/groenland-nummer-1-paa-facebook/). In Greenland's print media, Facebook is mentioned as a social medium that has a special status for Greenlanders, and everyone in the country knows what it is (Sermitsiaq, 2017; Sermitsiaq, 2016).

The major role played by social networks such as Facebook in Greenland and other Arctic communities is made possible "[as] mobile technologies become more affordable [and] engineers at companies such as Google and Facebook are optimizing web platforms through data compression technology for better use on low-bandwidth connections" (Yunes, 2016, p 99). Through its optimization for low-bandwidth connections, Arctic users use these platforms as means of intergenerational communication, advocacy and storytelling. Thus, Yunes describes how in the Arctic, "Social networking has become a mechanism for Arctic communities to reframe the conversation on Inuit life and culture in an increasingly powerful way." (2016).

Behind the increasingly impactful uses of social media in the Arctic are also local infrastructural changes such as increased access to affordable internet, which has greatly improved in the last decade. In Greenland, internet access is available to more than 90% of the population and is provided by the self-rule owned monopoly telecompany Tele Greenland A/S. However, the foundation for internet access varies in Greenland, which can be broadly divided into two zones: The 'Submarine and Microwave zone' covers around 87% of the population with a diverse infrastructure which primarily consists of submarine cables, such as Greenland Connect and Greenland Connect North (since 2017) and high-speed wireless links that expand the reach of the cables. The sea cables run across the south west coast, covering the main cities: Nuuk, Sisimiut, Ilulissat, and connect with Canada and Iceland (https://telepost.gl/en/telecommunications-infrastructure). Private Internet access in this zone offers up to 30 Mbps. In contrast, the 'Satellite zone' includes around 9% of the population living in the east and north west coasts. Cities in this zone include Tasiilaq, Ittoqqortoormiit, and Qaanaaq, where only up to 4 Mbps is offered.

Internet access is thus increasingly available in Greenland, but prices are considerably higher when compared to other countries. For example, the monthly subscription with the highest data rate, namely 30 Mbps, in Nuuk costs 899 DKK (around 120 €) per month (https://tusass.gl/private/internet, Accessed: 11 June, 2021). In contrast, users in remote areas must pay a higher price for a service that is considerably slower, as Tele Greenland does not compensate users who are more expensive to serve. For example, a 4 Mbps subscription in Tasiilaq costs 999 DKK (around 134 €) per month. These prices are considerably higher than in European countries, where an expensive fixed broadband subscription with 30 - 100 Mbps costs about 30.2 € per month in 2019 (European Commission, 2020).

Despite the widespread Internet coverage (WBG, 2019), the high subscription cost in combination with other socio-economic factors results in only 68.5% of the population in Greenland using the Internet in 2016 and only 17.74% having broadband connection (>256 kbps) in 2015. In contrast, the Broadband Europe initiative aimed to provide coverage for 30 Mbps or more for all citizens by 2020.

In addition to the increased cost and the comparatively low data rates, keeping Greenland's connectivity infrastructure functioning requires diligent maintenance and upkeep. This is reflected in frequent updates from Tele Greenland that describe how storms, solar interference, cable cuts from fishing trawlers and other environmental and human factors, which slow or completely shut down access to electricity, internet, radio or tv signals in settlements and towns across Greenland on a daily basis (see https://telepost.info)

In December of 2020, Tele Greenland posted an image on Facebook of the Greenland Connect and Greenland Connect North cables, with the caption "Let us together take care of all of Greenland's sea cable", where they underscored the importance of maintaining a safe distance to the undersea cables when sailing to avoid damage (https://www.facebook.com/telepostgl/videos/305914483456099). The post, however, was met with a user drily commenting that "It's like that cable does not cover 'all of' Greenland". This example illustrates how connectivity in a Greenlandic context is characterized by unpredictability as well concerns about differences - between urban and rural areas, those with high speed accounts and those not, those served by the cable or by satellite, and those affected by an interruption and those not. In this lens, experiences with connectivity in Greenland mirror those found in Australia by anthropologist Melissa Gregg, who identified a feeling of 'rural melancholia' among citizens in rural areas whose experiences in 'metro-centric' internet advertising were not depicted as ordinary: 'the feeling that manifests in thereflex (...) ofreading the fine print in advertising that says "availableinselectedmetroonly"' (Gregg 2010: 161) - or in the case of Greenland, sea cable access available! (in selected towns only).

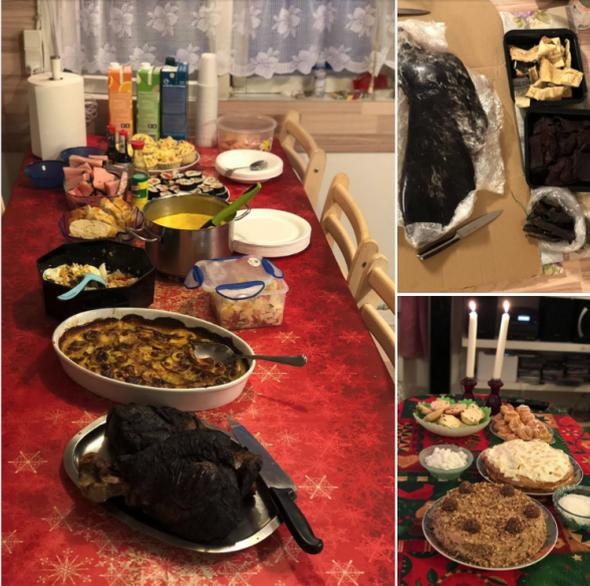

Figure 1: Post in the Ukkusissat allangarsiivia group from december 2020, shown here with the poster's permission.

Still, across the country, Greenlanders are taking advantage of the relatively low-bandwidth threshold for access to Facebook to share daily life with each other. Thus, use of social media leads to quick sharing of news and images within and across otherwise physically disconnected communities. In a country with high consumption of traditional mass media (e.g., more than five hours of radio every day), but low and diminishing production of Greenlandic content (Ravn-Højgaard et al., 2018), Facebook offers an alternative online news channel with access to vast amounts of local and global information.

Use of Facebook in Greenland is not in itself unusual or remarkable, unless we consider that Facebook in this way becomes a frugal infrastructure that connects citizens in a country where travel is expensive, taking place via plane or boat, as no towns or settlements are connected by a road. An example to showcase the practices and extent of online mundane connectivity is Ukkusissat, a settlement of about 150 inhabitants in the sparsely populated North-West Greenland. In the active group (an average of 8 daily posts according to Facebook) Ukkusissat allangarsiivia, or 'Ukkusissat messageboard', postings can be seen announcing a new baby, local news, things for sale, community invitations or updates such as the one showcased in Figure 1 depicting a recent 'kaffemik' (Greenlandic open house with food) celebrating a birthday, with many people commenting 'pilluaritsi', greenlandic for 'congratulations'.

The most interesting part of the message board for Ukkusissat is that it is followed by over 1800 members (per 29/6/2021), more than 10 times the number of current inhabitants in Ukkusissat. In fact, this pattern is repeated for message boards across Greenland's settlements and towns, where an audience many times larger than the local population follows and interacts with online posts, invitations and discussions.

These large followings for local channels of talk creates a way of bridging physical distances. Underlying this kind of use is the fact that Greenlanders born in settlements have to move away from their place of birth to seek education above elementary school-level, and many never return. As a result of the centralization earlier mentioned, there is a large percentage of Greenlanders who live far from their place of birth and are interested in following along and participating in township life. Online social networks allow former inhabitants, family and others who are interested in what happens in a particular settlement to follow along and partake in local events from a distance, in small frugal ways thus renegotiating national infrastructures.

These modest examples of situated, mundane connectivity demonstrate how community members use online platforms to cope with and overcome physical and social distance. This sensitizes us towards how communication infrastructure is a more-than-technological phenomenon and a site where connectivity - in its social sense - is strengthened but also at times challenged or even disabled. It offers an understanding of connectivity beyond merely being connected or disconnected, but rather as a spectrum of ongoing, locally negotiated, enabled and disabled connectivity. This challenges 'good infrastructure' as based on a one-size-fits-all model, in which the conditions of arctic connectivity needs drastic changing, but rather opens up for more adaptive, frugal models, which citizens and researchers can 'tinker' with.

**Fostering industrial innovation and global competitiveness in Greenland**

As discussed above, strengthened connectivity has already resulted in significant social as well as political change in the Arctic. Thus, access to fast and inexpensive broadband connection appears as an effective tool to reduce inequality among regions and, hence, to achieve SDG 10 (UN, 2015). However, increased connectivity and technology adoption also holds remarkable potential when considering industrial innovation and efficiency, in accordance with SDG 9: Build resilient infrastructure, promote inclusive and sustainable industrialization and foster innovation (UN, 2015). Therefore, it is one of the main objectives of numerous development plans such as the Broadband Europe and the 5G for Europe Action Plan (European Commission, 2016). Specifically, the worldwide race to deploy 5G is mainly driven by three concerns:

1. Ability to provide a dramatic increase in data rates and number of connections per unit area, which is termed enhanced mobile broadband (eMBB).
2. Ability to provide support for IoT applications and services. These can further be divided into two sets of applications (termed use cases) that aim to connect "things"
    1. Massive IoT connectivity: Targeting tens of thousands of devices exchanging small amounts of data in each cell/base station. Typical examples include small sensors and devices that are embedded in the environment and occasionally transmit small chunks of data: location, temperature, motion sensing, remote patient monitoring, etc...
    2. Ultra-reliable low-latency communications (URLLC): This refers to connections that offer a stable remote interaction among humans and machines. Examples include: remote control of a robot that works in a dangerous environment and remote surgery, etc. Differently from massive IoT, here URLLC targets extremely high guarantees that the data will be delivered within a given delay/latency deadline that does not exceed a few milliseconds.
3. Focus on vertical industries, providing a global platform for industrial innovation.

The benefits of broadband services have already been discussed in the latter section. We therefore focus now our attention towards the two other disruptive features of 5G in the context of Greenland: its support for IoT applications and for vertical industries. In this part, we discuss representative IoT applications in different vertical sectors that are relevant for the Greenlandic context. Unless explicitly stated, we will use the term IoT to denote 'massive IoT', while we will explicitly refer to IoT with ultra-reliable low-latency communication (URLLC) requirements.

*Healthcare*

Ensuring healthy lives and promoting well-being is one of the main points in the agenda for sustainable development, as reflected in SDG 3: Good health and well-being (UN, 2015). Furthermore, an efficient healthcare system is one of the main characteristics of strong economies and societies. One of the major use cases for IoT in e-health is remote patient monitoring, where small IoT devices or *wearables* sample and transmit the patients' health indicators. In general, the remote monitoring of patients enables safe and effective universal health coverage and risk reduction and prevention, which not only saves lives and reduces cost, but also facilitates the provision of personalized health care. In Greenland, the relevance of such applications is exacerbated by the long distance between settlements, which present a potential health risk for patients that need to reach a hospital with the required equipment during a medical emergency. For example, Tasiilaq is the seventh largest city in Greenland and the most populous in the East coast. A trip to the capital city, Nuuk, from Tasiilaq involves an aerial route of around 680 km. By performing an integral remote monitoring of the health of patients, such emergency trips can be minimized.

Asset tracking and monitoring is also relevant for the healthcare sector. A clear example is tracking the position and status (i.e., temperature, humidity) of vaccines and other medications throughout the supply chain. In the case of COVID-19, ensuring safe transportation of the vaccines is essential to overcome the pandemic. This tracking and monitoring can be achieved by attaching a set of IoT sensors to the assets of interest (i.e., wearables in the case of a patient being transported), but also requires continuous Internet coverage along the route.

Remote telemedicine is yet another interesting set of applications due to similar reasons as remote monitoring. These include, for example, telesurgery, where a surgeon operates a patient that is in a different location and tele-rehabilitation. Achieving telesurgery is a major milestone of the so-called tactile Internet (Simsek, 2016), as sensory and tactile feedback must be transmitted with ultra-low latency and ultra-high reliability is needed for the surgery to be performed successfully. Hence, this is a clear example of URLLC applications.

Recently, it was observed that 5G provides sufficiently high data rates and low latency for patient track and tracing and video transmission for telemedicine applications (Jell, et al., 2019).

*Fishing*

Fishing is the main industrial activity in Greenland, with shrimp, halibut and cod being the most commercial resources. A classic example of the impact of communication technology in fishing was provided by Jensen (2007) in a study of sardine fishermen and wholesalers in India. In the latter, it was found that the introduction of mobile phones to share price information increased the profits of the fishermen by 8%. This was mainly due to a reduced price dispersion and waste in the sardine catch.

The above example highlights how simple mobile Internet access impacts the fishing industry. However, companies and ports worldwide are adopting more advanced systems that enable process automation and real-time information exchange. In connected ports, machinery can perform automated tasks while being aware of its environment by utilizing a large number of IoT sensors to gain context awareness. Doing so maximizes efficiency and minimizes risks for the employees. In addition, deploying a communications infrastructure near the coastline in Greenland would enable fishermen and their vessels to communicate with each other as well as with nearby fish processing facilities to act as a coordinated unit. For example, fishermen would be able to use real-time mapping applications to show the position of the other vessels and their catch and to allow the vessels to coordinate with the portuary authorities to unload the catch with maximum efficiency. Besides, introducing such technologies facilitates accountability and transparency, so that the authorities can detect malicious behavior and enforce the regulations to conserve the marine life of the region, as covered by SDG 14: Conserve and sustainably use the oceans, seas and marine resources for sustainable development (UN, 2015)

*Mining*

Mining in Greenland is considered an essential activity to diversify its economy, which, as mentioned above, is greatly based in the fishing industry. Hence, Greenland's Mineral Resources Authority has developed a Mineral Strategy 2020-2024 to foster the growth of this sector (https://govmin.gl/2021/05/greenland-says-yes-to-mining-but-no-to-uranium/). In connection with the latter government initiative, the main mineral resources in Greenland, namely gold, zinc, nickel, diamonds, and rubies, have been mapped and their location has been made freely available (http://www.greenmin.gl/).

IoT applications for mining can be divided into the tools and techniques to explore and extract the minerals and the means for transportation, storage, accounting and to ensure sustainable industrialization (SDG 9). One of the major applications of IoT and the tactile Internet is robot telecontrol. The latter is an incredibly valuable asset in the mining industry, which enables the use of robots instead of human miners to perform the excavations in hazardous conditions to minimize the risks for the personnel. Naturally, an appropriate control of such robots presents similar, yet slightly more relaxed, URLLC requirements to telesurgery: in both cases sensory feedback (i.e., video, audio and touch) and tele commands are required to be delivered within strict deadlines.

Environmental monitoring is essential in the mining industry to ensure that the environmental impact and the pollutants due to mineral extraction and treatment are kept within the state regulations.

While the above-described applications appear to be widely different, they share the general features of IoT applications that distinguish them from human-to-human communication. The two most important features are 1) the transmission of relatively short amounts of data, in the order of a few tens of bytes up to a few kilobytes, and 2) the large number of communicating devices. These features render the communication protocols used for human-to-human communication ineffective.

Further, these features magnify the importance of reducing both the deployment (CAPEX) and operational (OPEX) cost per device. In particular, IoT devices would only be adopted by the industry if both, the devices themselves and the subscriptions to the wireless service were inexpensive. Hence, companies worldwide are commercializing low-cost tracking devices and around 158 cellular networks are offering IoT connectivity around the globe (https://www.gsma.com/iot/mobile-iot-commercial-launches/. Accessed: July 2, 2021). For example, in Europe, relatively simple IoT devices can be purchased from 39€ per device with 2-year service subscription included (https://iot.telekom.com/en/solutions/low-cost-tracker. Accessed: july 2, 2021.Alternatively, SIM cards for IoT devices are offered with a 60-month subscription from 7,95€

(https://iot.telekom.com/en/networks-tariffs/iot-tariffs/business-smart-connect. Accessed: July 2, 2021).

Another distinctive feature of IoT applications is the need to minimize energy consumption and maximize battery lifetime: replacing the batteries may be extremely expensive or impossible, so IoT devices are expected to operate for at least 2 years (and ideally, more than 10 years) with the same set of batteries. Hence, having the ability to go into any sort of power saving mode is greatly beneficial for IoT devices. On the downside, this hinders the communication with the base stations, and therefore the global Internet, for long periods.

To efficiently support IoT applications, 5G has been designed to provide greater flexibility in the radio-resource allocation when compared to 4G (3GPP TS 38.300), allowing efficient integration of sporadic transmissions of small data chunks and supporting low-power operation of IoT devices.

There are generally two types of IoT systems based on the spectrum in which they operate:
1) In a licensed spectrum, the mobile operator needs to buy an (expensive) license to have an exclusive right to use the spectrum within a given country or territory. This gives assurance that no other entities will create interference and thus decrease the quality of the wireless connections in that spectrum. All mobile communications (4G, 3G, 2G) are offered through a licensed spectrum. Hence, for the existing mobile operators the use of licensed spectrum is a natural choice for IoT as well, pointing towards technologies such as NB-IoT and LTE-M. Currently, NB-IoT has a great momentum among the mobile operators due to its extended coverage and reduced energy consumption.

2) In an unlicensed spectrum, the operator does not need to have a license, but the deployed devices and infrastructure need to respect certain rules of "politeness" when using the spectrum. A typical example is Wi-Fi: anyone can buy and install as many Wi-Fi access points as desired, but, even though these devices follow the politeness rules, the communication performance degrades when many Wi-Fi transmitters are concentrated in a small area. Nevertheless, considering that IoT devices usually generate small amounts of data sparingly (i.e., sporadic activation), they tend to occupy a small fraction of the wireless resources for communication and unlicensed spectrum can be a viable option. Two widespread IoT technologies in unlicensed spectrum are LoRa and SigFox (Qadir et al., 2018). In terms of the politeness rule, LoRa and Sigfox devices follow duty cycling:

transmit only a pre-specified and small fraction of the time. While operation in an unlicensed spectrum is prone to interference, it lowers the barrier for new IoT operators and fosters innovation in terms of infrastructure.

Since IoT applications may have greatly diverse characteristics and requirements in terms of, for example, latency, reliability, battery lifetime, coverage, mobility, and cost, there is not a single answer to the question of which IoT technology is best. However, having a flexible communication infrastructure, into which diverse IoT technologies can be integrated depending on the localized connectivity needs, seems to be the optimal choice.

In the following section, we examine which novel technology enablers could develop frugal connectivity infrastructures in Greenland.

**Frugal and collaborative infrastructures for the Greenlandic landscape**

As previously discussed, the upcoming 5G system has positioned itself as a disruptive technology that will bring numerous benefits to communities that are already well-served by the current 4G, termed digital oases (Saarnisaari et al., 2020). For remote communities and for devices outside of these digital oases, however, it will offer little to no new advances towards increased coverage - not at least in the first phase of deployment as currently planned.

Presently, as outlined by Yaacoub and Alouini (2020), there is no single technology that is able to provide efficient broadband Internet access to remote communities and, at the same time, support the wide range of IoT applications that may arise. Hence, to serve an environment such as that of Greenland, we therefore advocate the design of flexible communication infrastructures that can be easily extended using a diverse set of technologies. In particular, we advocate the use of diverse communication technologies in: 1) the backhaul links, connecting the infrastructure, such as base stations (BSs), satellites, etc., to the Internet; 2) the fronthaul links, the (optional) links connecting part of the infrastructure with limited functionalities to fully-functional infrastructure with control units, such as BSs; and 3) the access links, connecting the end users to the infrastructure (Yaacoub and Alouini, 2020). Figure 2 illustrates the backhaul, fronthaul, and access links in a complex infrastructure with fully-functional BS (with direct Internet access) and BS with limited functionalities (with indirect Internet access), satellites, satellite ground stations, drones. Distinct types of user terminals (end users), namely, IoT devices, smartphones, industrial complexes (e.g., factories), ships, and households are also illustrated.

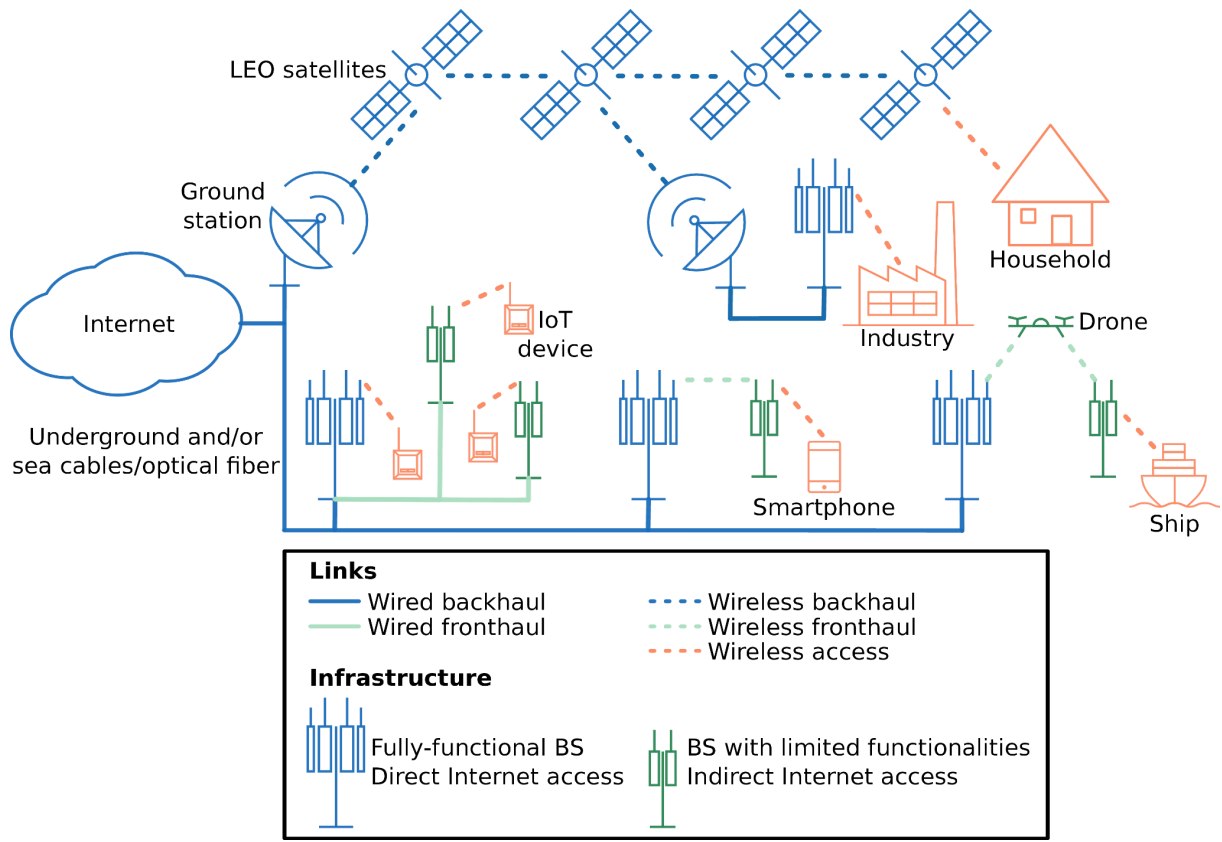

Figure 2: Example of backhaul, fronthaul, and access links in a complex infrastructure with fully functional BSs (with direct Internet access), BSs with limited functionalities (with indirect Internet access), low Earth orbit (LEO) satellites, and ground stations (to connect the terrestrial and satellite infrastructure). End users include IoT devices, smartphones, households, industrial facilities, and ships.

The flexibility provided by such an infrastructure would enable a collaborative approach to infrastructural development, where prioritization and choice of communication technologies can be led by local communities, rather than national or global concerns.

Three novel technologies/paradigms emerging in 5G context that could allow for the envisioned flexibility are: network function virtualization (NFV), software defined networking (SDN) (Akyildiz, 2015)., and Open Radio Access Network (Open RAN) architectures (ORAN Alliance; Wand and Kelly, 2019). The tendency in all of them is democratization of the innovation and ownership of the telecom infrastructure, relying on reconfiguration in software rather than dedicated, expensive hardware elements. Specifically, Open RAN aims to interconnect pieces of network hardware by defining and implementing open interfaces in software, enabling seamless interworking among commercial and all-purpose equipment from different vendors at the network nodes. SDN, on the other hand, separates the control and user planes, enables the use of flexible and centralized control units, and abstracts the network elements. Finally, NFV abstracts the network functions so these can be implemented in software. By doing so, instead of requiring dedicated hardware elements, the network functions can be easily configured and optimized in software running in generic hardware platforms. In effect, this approach makes the network 1) easily configurable, by executing centralized control commands to reconfigure the network functions (SDN); and 2) frugal and flexible, by selecting and deploying cost-efficient all-purpose processing equipment and antennas based on the localized connectivity needs, which in turn enables the integration of distinct radio access technologies and the provision of performance guarantees to a wide range of services with heterogeneous requirements. Figure 3 provides an example of a flexible network architecture based on Open RAN compared to a traditional network architecture, which can reduce deployment costs due to the ability to integrate equipment from different vendors.

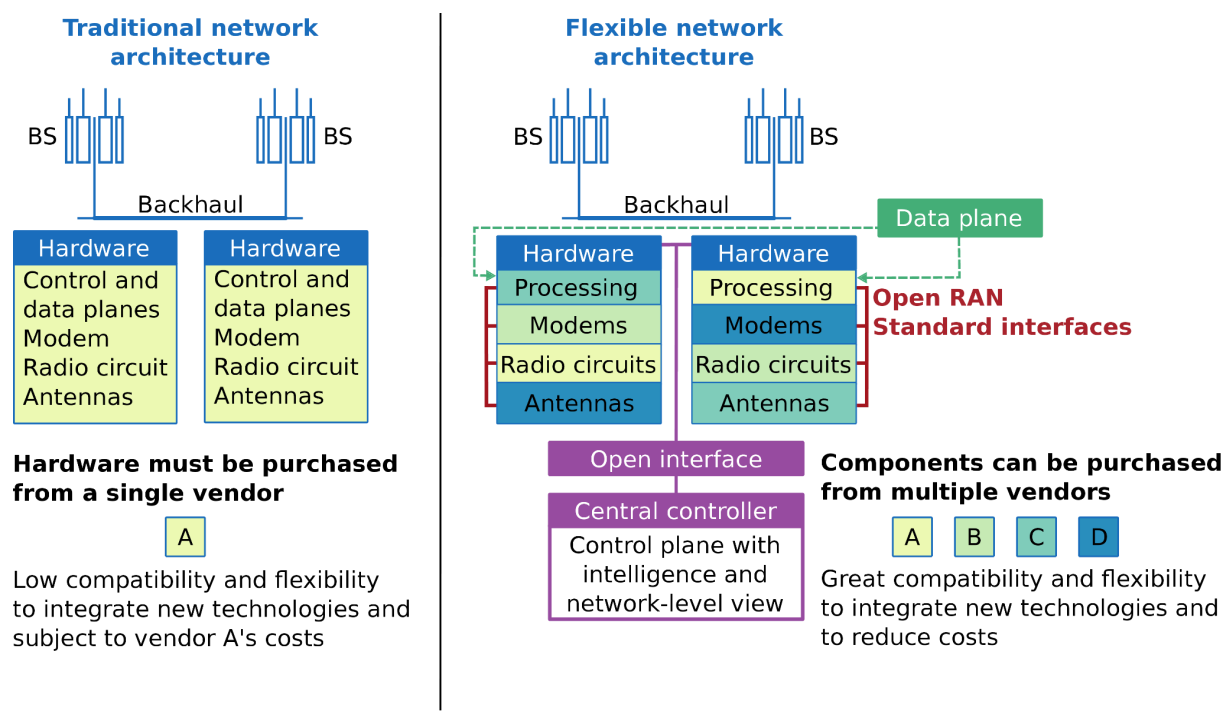

Figure 3: Comparison of a traditional network architecture with a flexible network architecture based in Open RAN. The latter enables the implementation of network functions

in software and the use of components from different vendors and technologies, which can reduce costs while fulfilling the specific needs of the end users within coverage.

**Future frugal connectivity architecture**

In the following, we investigate technologies that can be integrated into a frugal Open RAN architecture. We divide these technologies based on their role as either in the backhaul, the fronthaul, and in the access network.

Satellite networks may be the only option that is economically viable to provide a reliable fronthaul and backhaul, solving the interruption of connections due to cable cuttings and sometimes offering the only option to provide coverage The use of satellites for backhaul is not, by itself, novel and the following three main business models for mobile network operators (MNOs) offering cellular service through satellite backhaul links have been developed (ESOA, 2016).

1. The MNO contracts the satellite service and buys a ground terminal, hence, it manages its own service.
2. The MNO contracts the satellite service and the ground infrastructure, so that the satellite operator manages the satellite and ground infrastructure.
3. The MNO enters an agreement with a service provider to use its end-to-end connectivity solution.

Currently, the infrastructure in Greenland incorporates large satellites in high Geostationary orbits (GSO), but recent advances in the satellite industry regarding satellites deployed at low Earth orbits (LEO) are opening the door to new and interesting possibilities. Such satellites can offer lower latency in the data transmission than those at GSO due to the shorter transmission paths but require a densely deployed infrastructure.

With the aim to provide ubiquitous 5G coverage, the 3rd Generation Partnership Project (3GPP) that is in charge of the 5G standardization has started the activities towards integrating satellites into 5G (3GPP 2018, 2019). The initial Release 17 considerations concern the use of LEO satellites as simple relays towards a base station located at ground level. In addition, the efforts to integrate satellites into 5G are progressing slowly and may not be concluded until the release of 6G (Saarnisaari et al., 2020) by 2030. This long-term perspective is currently insufficient, as it would not fulfill the ambitious goals of the

Broadband Europe initiative and of SDG 9 (UN, 2015), which aim to increase the percentage of the population with access to mobile broadband connectivity and to provide at least 100 Mbps in rural areas by 2025. Nevertheless, numerous private companies such as SpaceX, Telesat, OneWeb, Amazon (project Kuiper), and Kepler, are currently investing in and deploying networks of hundreds and thousands of satellites at LEO. While these will not be integrated directly into 5G, the development of open interfaces under the Open RAN paradigm would provide an efficient solution. However, the design of commercial satellite constellations is driven by the global business opportunities and, hence, the end goal is to provide global connectivity to most of the Earth's population, of which the Arctic population is only a small fraction. For instance, Starlink's satellites deployed at 550 km above the Earth's surface have the potential to provide low-latency connectivity (Handley, 2019). Nevertheless, these satellites cannot provide coverage in Greenland. Specifically, they are organized in a type of constellation called Walker delta, and only cover latitudes of up to 53°, whereas the latitude of the southernmost city in Greenland, Aappilattoq, is 60°.

Kepler satellites, on the other hand, are organized in a different type of constellation called Walker star where the satellites orbit near the poles. Hence, Kepler's constellation is able to provide IoT service in the polar regions. However, Kepler's satellites follow a store-and-forward approach, where the satellites receive the data and store it until they move within the coverage of a ground station. In effect, the reported data experiences rather large delays — in the order of minutes — which may be restrictive even for a wide range of IoT applications that support delays of up to a few seconds. It has been emphasized that communication between satellites through free-space optical and/or radio frequency (RF) links are essential to achieve a fully functional satellite backhaul (Soret, 2021) and to significantly reduce the delay of data transmission in satellite constellations, but there are still several challenges, mainly due to the rapid movement of the satellites and their limited capabilities, to overcome to achieve efficient inter-satellite communications.

In combination with a satellite backhaul, free-space optical and/or long-range RF links can be used in the fronthaul, connecting several base stations with limited capabilities to a more capable central controller, to extend the coverage around communities served by the current (cabled) infrastructure or by the satellite backhaul. For instance, satellites acting as relays could connect a remote base station with minimal functionalities to a fully functional base station connected to the Internet. Besides,the current infrastructure already incorporates terrestrial high-speed RF fronthaul links, but the use of open interfaces via Open RAN can greatly reduce their cost, enable the integration of different technologies, and enable the reuse of existing infrastructure, such as television towers. For example, the Frugal 5G network architecture (Khaturia, Jha and Karandikar, 2020) tries to "connect the unconnected" by combining a set of components from the evolving 5G cellular standards along with wireless local area networks (WLAN) operating in unlicensed spectrum. While this is a cost-effective solution, the use of 5G in the core network, Ethernet for the fronthaul, and WLAN for the access network greatly limits the mobility and the number of supported users. Hence, it may result in a great difference between the experienced QoS at different points in the infrastructure. This goes against the objective of providing comparable QoS guarantees for broadband services in remote and densely populated areas. However, it may be suitable to provide downscaled versions of essential IoT applications.

An interesting idea to provide geographically targeted solutions in the access network is the use of unmanned aerial vehicles (UAV) such as balloons and/or drones. These may even be able to communicate with satellites to form a multi-layered network. Telelift, for example, employs tethered drones that serve as LTE (i.e., 4G) base stations. These drones solve the energy supply problem by being attached to a battery or power outlet on the ground and can fly for extended periods. Given the appropriate interfaces are implemented as part of an Open RAN architecture, these drones could be connected to the core network through a dedicated satellite backhaul. Finally, as indicated by Yaacoub and Alouini (2020), some applications do not explicitly require Internet connectivity and a simple local/community network can provide the required performance with a reduced cost. This is the case of, for example, communication between vessels, described in the applications relevant for the fishing industry. Here, the local traffic can be maintained within a local wireless mesh infrastructure and, hence, can be included in a low-cost subscription and kept independent from the traffic that must traverse the Internet.

**Conclusion: A future collaborative research agenda for arctic connectivity**

As discussed in the introduction, an implementation in the Arctic of the 5G system in its current phase of standardization risks repeating the existing macro- scale logics that have resulted in infrastructural solutions with little relevance for the Arctic populations themselves (Schweitzer and Povoroznyuk 2019). In contrast, we emphasize that infrastructure is a more-than-industrial concern, entangling with other Arctic everyday concerns. Maintaining a business, keeping in touch with friends and family, receiving proper health care and attending school are all dependent on local stable, relevant infrastructures. Infrastructure is also tied up with notions of access and difference, as infrastructural development reflects resource allocation and on the most basic level works to support or undermine the development of Arctic communities. Our focus on the social aspects of infrastructure and the importance of community involvement echoes the finding by Pirinen et al (2019) that technology itself is not the main hindrance to improvements in Arctic connectivity. Across the Arctic, hindrances rather take the form of lack of commercial/economical incentive as well as political interest in prioritizing remote areas.

In this article, we have critically challenged the standardized 5G research agenda and proposed a new way to approach infrastructural development, introducing the principle of 'frugality': A careful balance of local needs to make the most of existing possibilities while providing valuable and flexible solutions. We argue that no single technology will be able to provide efficient broadband Internet access to remote communities and, at the same time, support the wide range of IoT applications that may arise. Therefore, a frugal approach to arctic infrastructural development will entail collaborative work, where prioritization and choice of communication technologies is led by local communities. Communities, as exemplified by the use of Facebook in Greenland to cope with and overcome physical and social distance, who are already engaged in making the most of the available infrastructure.

However, while we have suggested examples of valuable and cost-effective connectivity practices in both civil society and industry, more research on potential use cases for frugal connectivity in an Arctic context is needed to identify possible stakeholders and technologies of interest. Most importantly, more research is needed on how frugal infrastructural development might be practically undertaken on a larger scale in arctic communities, in collaboration with both community members and local businesses.